# Agent Development Toolkits

**Aarti Singh [1], Dimple Juneja[1], A.K. Sharma[3]**
M.M.University,Mullana (Ambala), Haryana, India
Y.M.C.A University of Science and Technology, Faridabad , Haryana , India
singh2208@gmail.com

Development of agents as well as their wide usage requires good underlying infrastructure. Literature indicates scarcity of agent development tools in initial years of research which limited the exploitation of this beneficial technology. However, today a wide variety of tools are available, for developing robust infrastructure. This technical note provides a deep overview of such tools and contrasts features provided by them.

**A Technical Note**

## 1. Introduction

Large scale realization of agent applications requires frameworks, methodologies and tools that support effective development of agent systems. Historically the main obstacle in agent and multi agent system development used to be the infrastructure which refers to supporting environment where agents can communicate and achieve their desired goals. However, simply having an infrastructure is usually not enough. As user-acceptance of the infrastructure depends on ease of application development, versatility of applications, support for various standards & most importantly the security of communication among agents, security of agents on remote platforms and security of mobile agent hosting platforms.

Nowadays many agent development tools and platforms of different quality and maturity are available and have been employed for different applications in different parts of the world. There is so far no consensus about which tool is best for agent development.  With increased popularity of agent technology, agent development has also received attention, securing its place in commercial sector as well. Thus it is worth studying various agent toolkits in depth, and to analyze their strengths & weaknesses.

## 2. Agent Development Tools

According to the technical report by Nguyen & Dang [13] there are over 100 products in this category. Due to the space constraints, we would be focusing on most appealing *& promising toolkits among the available choices.

### *2.1   IBM-Aglet or Aglet Software Development Kit (ASDK)*

ASDK or IBM-Aglet [13, 8, 10] is an environment for developing mobile agents based application in JAVA. It is an open source freely available toolkit, with latest version Aglet 2.5 alpha. It provides good graphical user interface for agent development. It mainly comprises of two packages-The Aglet Building Environment (ABE) and the Aglet Workbench. Aglet workbench aims at developing stand alone mobile agents. The ABE (SDK) comprises of Aglet API, the Aglet Server known as Tahiti and the Agent Web Launcher called Fiji along with documentation and sample Aglets. Aglets are basically java objects comprising of two major components i.e. Aglet Core & Aglet Proxy. Core is holder of all the internal variables and methods of an agent whereas proxy acts as an interface to the core, shielding it from any malicious interference from the outside world.

Aglet server Tahiti is an application program that works as agent server for aglets. It provides users with a good GUI and allows users to create & dispatch an agent, monitor it, dispose it off when required. It gives user the ability to set agent's access privileges on the server. For an aglet to move to a remote host, it must have Tahiti server installed on it, which solves some of the security problems. Fiji is an applet in java which can create Aglets or retract an existing aglet into client's web browser. This applet accepts an agent's URL as parameter and can be embedded in a web page using HTML, like any other applet.

Aglets support both agent mobility as well as predefined movement of the agent on the network also called as Itinerary. Although aglet provides weak agent mobility but that too is restricted to its own servers. Aglet works on Mobile Agent System Interoperability Facility (MASIF). Agent migration is implemented using socket mechanism. Communication among agents is achieved using synchronous and asynchronous message passing. Agent Transfer Protocol (ATP) along with Java Agent Transfer and Communication Interface (J-ATCI) also help achieve the same

Although Aglet platform has wide user acceptance but it doesn't provide much security. Its security is knitted in the concept of restricting transfer of aglets only to its own servers. Due to lack of security, state of aglets can't be stored on any other host. No such method is provided by this tool. Scalability is another problem, since aglets are not interoperable with other platforms or their agents, due to their restriction of working with their own server. Following figure illustrates the structure of an aglet

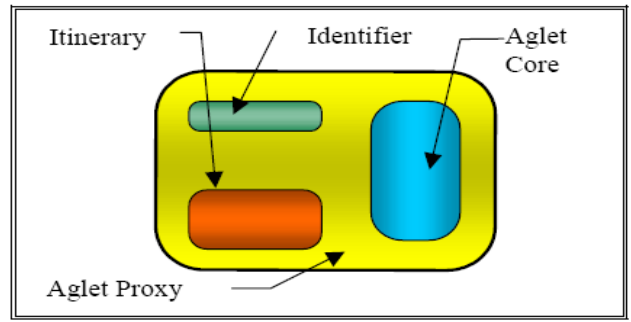

Fig. 1: Structure of an Aglet [8]

### 2.2 Voyager

Voyager [10, 1, 9 ] is an agent development tool developed by ObjectSpace, in mid-1996. ObjectSpace has been taken over by Recursion Software Inc. since 2001 and it's now their commercial product. Latest version available is Voyager 8.0. It's a simple yet powerful technology for creating mobile agents in Java. It was an improvement over already existing platforms like Aglets, Odyssey, Concordia etc. which only allowed developers to create agents and launch them into a network to fulfill its mission. But none allowed sending messages to a moving agent, which made it difficult to communicate with an agent once it has been launched and also for agents to communicate with other agents.

*A Technical Note*

Voyager seamlessly integrated fundamental distributed computing with agent technology. It treats an agent like a special kind of object which can move independently and can continue its execution while moving around. The agent and object are different only because an agent can move autonomously whereas an object can't.

It allows an agent to send and receive Java messages to and from other agent, even while traversing the network, irrespective of its position in the network. It supports synchronous, one-way and future message modes. Whenever an agent moves, it leaves behind a forwarder object which forwards the message to its new location.

Voyager provides flexible life spans for agents, by supporting a variety of life span methods:

*(1) An agent can live until it has none local or remote references (default life span of an agent).*
*(2) Agent can live for a certain amount of time (by default for a single day).*
*(3) It can live for a particular point in time.*
*(4) It can live until it remains inactive for a specified time.*
*(5) An agent can live forever.*
*(6) An agent's life span can be changed flexibly as required.*

Another attractive feature of this tool is its support for directory service, which is particularly important in launching a mobile agent from one application to another and for locating an agent after it moves to some other location in the network. Its directory structure allows creating and connecting network directories together to generate a large interlinked directory structure.

Voyager supports weak mobility of agents using RMI technique. It allows all serializable objects to be mobile using Virtual Code Compiler (VCC). VCC utility accepts any .class or .java file and produces a new remote enabled virtual class. This virtual class is used in further communications with that agent/object. Agents use moveTo() function and a callback function for migrating to a remote host. On reaching new host, the agent retrives the callback function that it sent and resumes its execution.

Voyager has an associated server called 'voyager' but it's not necessary to have such server installed on all nodes in the network. Due to this reason agents created using voyager are provided restricted access on the host servers. Thus provision of agent and host security is weak in this tool.

### *2.3 JADE*

JADE (Java Agent DEvelopment Framework) [13, 3, 4, 6 , 5 , 17] is a software Framework fully implemented in Java language. It is developed by Tilab for the development of multi-agent applications based on peer-to-peer communication architecture. Latest version available is 4.0.1 released in July 2010. It simplifies the implementation of multi-agent systems through a middle-ware that complies with the latest Foundation for intelligent physical agents (FIPA) 2000 specifications. It provides a set of graphical tools that supports the debugging and deployment phases of agent development. Jade permits the intelligence, information & resources to be distributed over the network in the form of java compatible mobile devices like PDA, pagers, cell phones, smart phones, laptops or fixed desktops etc. The communication environment evolves gradually with the appearance and disappearance of various peers, (known as agents in Jade) according to their needs and requirements.

In JADE an instance of run-time environment is called a container, as it holds all the agents created in it. Collection of such containers is called a platform and it provides a homogenous layer which hides the complexity of underlying hardware & software from agents and their developers. It is compatible with J2ME, J2EE & CLDC. Its low memory requirements make it suitable for mobile devices. Nokia, Motorola, Siemens, Compaq & Hp are some of well known brands having compatibility with JADE. Following figure illustrates the architecture of JADE.

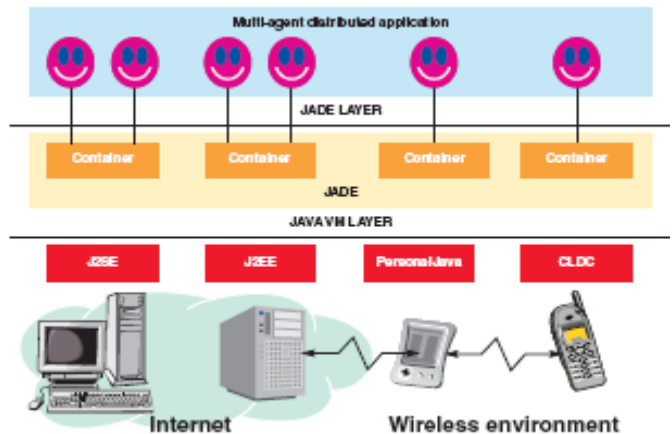

Fig.2: Architecture of JADE [ 3 ]

Every container in JADE comprises of Directory Facilitator (DF) agent, Remote Monitoring Interface (RMI) agent & Agent Monitoring System (AMS) agent. Here the agents can discover other agents dynamically using DF agent & can communicate with each other using peer-to-peer paradigm.

In JADE agents communicate using asynchronous message passing technique which is most widely accepted model for distributed and loosely coupled communications. JADE preserves the security of agents by providing strong authentication mechanisms to verify rights of any agent. Messages exchanged among agents comply with Agent Communication Language (ACL) defined by FIPA. JADE supports execution of multiple parallel tasks with in the same java thread. This feature ensures scalability as well as meets resource constraints of environment.

It supports agent mobility by allowing an agent to transfer its code as well as its state to remote hosts. The form of mobility supported in JADE is known as 'Not-so-Weak' as the stack & the program counter can not be saved in Java. Jade has security feature inbuilt in it. JADE Object Manager provides connection authentication, user validation and RPC message encryption. The Jade socket proxy agent acts as bidirectional gateway between a JADE platform and an ordinary TCP/IP connection [13].

Interoperability is an attractive feature of JADE, since it is FIPA compliant. Thus agents created in JADE can interoperate with other agents, following the same standards. Also it supports complex interaction protocols like contract net protocol for facilitating complex multi-agent applications.

*A Technical Note*

*A Technical Note*

### 2.4 Anchor

Anchor [12] agent toolkit is developed by Lawerence Berkeley National Laboratory, U.S.A. it facilitates the transmission and secure management of mobile agents in a heterogeneous distributed environments. This toolkit is available in BSD style license. Its architecture comprises of an agent viewer graphical user interface, Agent API, Anchor server, Anchor security manager (ASM) , Anchor class loader (ACL), secure agent transfer protocol (satp) handler, Anchor Java Naming and Directory Interface (AJNDI) & Anchor Java Native Interface (AJNI) components.

Agent model in Anchor is based on that of Aglets. Agents are serializable java objects capable of migrating in the network. Agents are created in contexts, where context is a namespace under which agents are grouped together. Agents are accessed through their proxies, which protect the agent from any attempt of direct access to its code and methods. Also it provides location transparency to agent which means an agent is represented in a machine even if it has migrated to some other machine in the network. All messages to that agent are forwarded to its location through its proxy.

Agent server in Anchor is a run time environment which acts as backbone of this toolkit. It runs on a host and works on a specific port. It performs all system related functions. Anchor server supports the agent migration through satp. Security is a major concern in this toolkit. Mutual authentication between agent systems is established through secure socket layer (SSL). Agents are authenticated by signing their byte codes with their private keys.

An interesting component of Anchor toolkit is Akenti which is an access control system designed to ensure controlled access of distributed resources. This component uses Public key infrastructure. Access control decisions are made using digitally signed certificated based on X.509 standard.

Integration of Akenti component in Anchor provides it strong security. AJANDI component in Anchor provides a naming service through which every agent can register and publish its current information. It also provides a directory service to enable effective searching of agents. Agent viewer component supports features like creating an agent, its dispatch, retraction, disposal, activation and deactivation and also cloning of an agent.

### 2.5 Zeus

Zeus [13, 7, 18] is an integrated environment for the rapid development of collaborative agent applications, developed by Advanced Applications & Technology Department of British Telecommunication labs (http://www.bt.lab.com). It is open source freely available toolkit. It is purely implemented in Java which makes it compatible with most hardware platforms. It also complies with FIPA standards.

Zeus provides support for generic agent functionality and has sophisticated support for the planning and scheduling of an agent's actions. It provides a set of software components and tools used to design, develop and organize agent systems. It has good graphical user interface and embedded components like report generation tool, statistical tool, agent and society viewer tool etc. which help in observation of application under development. Also it allows the designers to use different negotiation techniques for testing implemented agents.

Communication among agents is performed using Agent Communication Language (ACL) or Knowledge Query Manipulation language (KQML). Communication security is provided using public key, private key cryptography and digital signature technologies.

Major drawbacks of this toolkit include lack of support for agent mobility and its weak documentation which leads to difficulties in creation of new applications.

## 3. Comparison of Various Toolkits

This section compares the above toolkits on major features which can affect their applicability.

- *Nature of product*: whether a toolkit is open source or commercially available influences end users a great deal, as if a product is freely available some compromises can be made with features provided as compared to the one which involves some cost.
- *Standard Implemented*: Foundation for Intelligent Physical Agents (FIPA) is a non profit organization involved in standardization of protocols & specification for agents Its goal is to establish internationally accepted specifications so as to maximize interoperability across agent based applications, services and equipment. Thus if a toolkit comply with FIPA standards it increases its utility and scalability.
- *Communication Technique:* Asynchronous communication is more efficient as compared to synchronous communication.
- *Security Mechanism:* Strong security mechanisms are desirable in technologies operating in heterogeneous distributed environments like world wide web (WWW). Considering the nature of applications where agents are employed these days, strong security features embedded in a toolkit can make it more appealing.
- *Agent mobility:* agent mobility can reduce network traffic and can increase efficiency of agents. It's a feature desired from agent toolkit.
- *Migration Mechanism:* RMI mechanism consumes more time and resources in agent migration compared to socket mechanism.

Table 1 given below summarizes the features of various toolkits discussed so far in this paper.

**Table 1: Comparison of Various Toolkits**

| Agent Development Toolkits → <br> Features ↓ | Aglet | Voyager | JADE | Anchor | Zeus |
|---|---|---|---|---|---|
| *Nature of Produce* | Free, Open source | Commercial | Free, Open Source | Available in BSD license | Free, open source |
| *Standard implemented* | MASIF | ---- | FIPA Compliant | SSL, X.509 | FIPA compliant |
| *Communication Technique* | Synchronous , Asynchronous | All methods | Asynchronous | Asynchronous | Asynchronous |
| *Security Mechanism* | Poor | Weak | Good | Strong security | Good |
| *Agent Mobility* | Weak | Weak | Not-so-weak | Weak | Do not support |
| *Agent Migration Mechanism* | Socket | RMI | RMI | Socket | null |

## 4. Conclusions

This work analyzed five agent development toolkits developed by different groups. On comparison Jade agent development toolkit seems most appealing. It is open source platform, purely designed in Java, provides consistency in API and supports different kinds of devices operating in internet. It provides good security features and supports sound agent mobility. Among other tools Voyager is commercial tool and doesn't comply with FIPA standards. Zeus supports FIPA standards but doesn't provide agent mobility. Aglet also doesn't comply with FIPA, lacks security and scalability. Anchor provides good security but doesn't follow FIPA specifications, thus lacks scalability. Thus JADE agent development toolkit is most balanced toolkit among the five discussed in this work.

A Technical Note